\documentclass[prl,preprint,showpacs,amsmath,amssymb]{revtex4}
\usepackage[utf8x]{inputenc}
\usepackage{graphicx}
\usepackage{float}
\usepackage{dcolumn}
\usepackage{bm}
\usepackage{sidecap} 
\usepackage{amssymb}
\usepackage{epsfig}
\usepackage{psfrag}
\def\beq{\begin{equation}}
\def\eeq{\end{equation}}

\begin{document}
\title{Sharp Raman anomalies and broken adiabaticity at a pressure induced transition from band to topological insulator in Sb$_2$Se$_3$}
\author{Achintya Bera$^1$}
\author{Koushik Pal$^{2,3}$}
\author{D. V. S. Muthu$^1$}
\author{Somaditya Sen$^4$}
\author{Prasenjit Guptasarma$^4$}
\author{U. V. Waghmare$^3$}
\author{{A. K. Sood$^1$\footnote[1]{electronic mail:asood@physics.iisc.ernet.in}}}

\affiliation{$^1$Department of Physics, Indian Institute of Science, Bangalore-560012, India}
\affiliation{$^2$ Chemistry and Physics of Materials Unit, Jawaharlal Nehru Centre for Advanced Scientific Research, Bangalore-560064, India}
\affiliation{$^3$ Theoretical Sciences Unit, Jawaharlal Nehru Centre for Advanced Scientific Research, Bangalore-560064, India}
\affiliation{$^4$ Department of Physics, University of Wisconsin-Milwaukee, Wisconsin-53211, USA}

\begin{abstract}

The non-trivial electronic topology of a topological insulator is so far known to display signatures  in a robust metallic state at the {\it surface}. Here, we establish vibrational anomalies in Raman spectra of the {\it bulk} that signify changes in electronic topology: an E$^2_g$  phonon softens unusually and its line-width exhibits an asymmetric peak at the  pressure induced electronic topological transition (ETT) in Sb$_2$Se$_3$ crystal.   Our first-principles calculations confirm the electronic transition from band to topological insulating state with reversal of parity of electronic bands passing through a metallic state at the ETT, but do {\it not} capture the phonon anomalies which involve breakdown of adiabatic approximation due to strongly coupled dynamics of phonons and electrons. Treating this within a four-band model of topological insulators, we elucidate how non-adiabatic renormalization of phonons constitutes readily measurable {\it bulk} signatures of an ETT, which will facilitate efforts to develop topological insulators by modifying a band insulator.

\end{abstract}

\maketitle


Topological insulators (TIs) are a new class of materials which exhibit an electronic band gap and a topologically nontrivial electronic structure in their {\it bulk} form \cite{ moore1,hasan1,model1,mele1,fu1}, of which the latter has interesting consequences for electronic states on their surfaces;  for example, one-dimensional edge spin states in 2-dimensional quantum spin Hall systems. A 3-dimensional TI with time reversal symmetry \cite{hasan1,zhang1,inversion1,tis1,quasi1,tight1,dirac1,fpri1} exhibits gapless surface states. While the transport properties  of a TI are theoretically predicted to be influenced by the topology of its electronic structure \cite{culcer1}, much of its experimental confirmation comes from studies of the surface electronic structure \cite{hasan1}, specifically on its odd number of Dirac cones. The nontrivial topology of electronic structure  results in an intrinsic magneto-electric coupling \cite{magneto2}, yet to be explored experimentally. Although the signatures of the electronic topology in the {\it bulk} properties may be subtle, they are expected to be more readily detectable when there is a sharp change in the electronic topology, {\it i.e.} at an electronic topological transition.

Electronic structure of a TI is characterized by topological invariants \cite{fu1,bal1}, which are determined from geometric properties of electronic states as a function of the Bloch vector. Similar ideas are involved in the theory of electric polarization \cite{berry1}, which is determined as a geometric phase of Bloch states. An integer quantum of change in polarization has been shown to arise from a cyclic evolution (adiabatic pumping) of the insulator along a path that encloses a metallic state \cite{uv1, rabe1}. Similarly, a bulk state with vanishing electronic gap in the vicinity of a TI influences the geometric properties of electronic states and is relevant to an electronic topological transition (ETT) \cite{naga1}. It will be interesting to probe effects of the resulting slow dynamics of electrons at the ETT, particularly on phonons which would require going beyond the adiabatic approximation.

An occurrence of broken adiabaticity has been seen in graphene, also characterized by a vanishing band-gap:  explanation of vibrational signatures of doping in graphene probed by Raman spectroscopy required going beyond the  Born-Oppenheimer (adiabatic) approximation \cite{ferrari1,andy1}. Similar nonadiabatic effects arising from the vicinity of a TI to a metallic state prompt us to carefully look for subtle reflections of an ETT in Raman spectra.

Among the 3-dimensional time-reversal invariant topological insulators, group V-selenides and tellurides (Bi$_2$Te$_3$, Bi$_2$Se$_3$ and Sb$_2$Te$_3$) have been extensively studied \cite{zhang1,poli1,bs0,gopal1,bs1,bs2,chao1}. In this family of compounds, the spin-orbit coupling (SOC) competes with crystal-field splitting and chemical  bonding (hybridization) between cation and chalcogen, and is strong enough to give rise to a topological insulator characterized by an overall odd parity of the valence band manifold \cite{zhang1}. Vilaplana et al \cite{bs1} examined an ETT in Bi$_2$Se$_3$ as a function of pressure using Raman spectroscopy and calculations, though the nature of changes in electronic structure or topology needs further clarification. Interestingly, a close cousin of these compounds, Sb$_2$Se$_3$ is a band insulator with an overall even parity of the valence band manifold, in contrast to an overall odd parity exhibited by the TIs (e.g. Bi$_2$Te$_3$) \cite{zhang1}. Under uniaxial strain, Sb$_2$Se$_3$ has been shown theoretically to undergo a transition to a TI state \cite{zhong1}. Thus, Sb$_2$Se$_3$ is an ideal host system for exploration of a possible ETT and its reflections in bulk properties. Competition between the spin-orbit coupling and hybridization can be tuned to introduce an ETT either by substitution of Bi at the Sb site or by application of pressure. While the former would alter the SOC, the latter would tune the crystal field splitting and hybridization between Sb and Se. 

Here, we use a combination of Raman experiments on single crystal Sb$_2$Se$_3$ and first-principles density functional theory (DFT) of electronic structure to establish the presence of an ETT in Sb$_2$Se$_3$ as a function of pressure.  As these calculations do not capture the observed phonon anomalies at the ETT, we derive symmetry invariant forms of electron-phonon coupling, and go beyond the adiabatic approximation using a model Hamiltonian in our analysis, and uncover mechanisms responsible for the anomalies in Raman spectra of Sb$_2$Se$_3$ signifying the change in electronic topology at its ETT. In particular, the asymmetry in observed phonon anomalies at the ETT arises from reversal of parities of electronic states across the gap.


\begin{figure}[p!]
\begin{center}
\leavevmode
\includegraphics[width=1.0\textwidth]{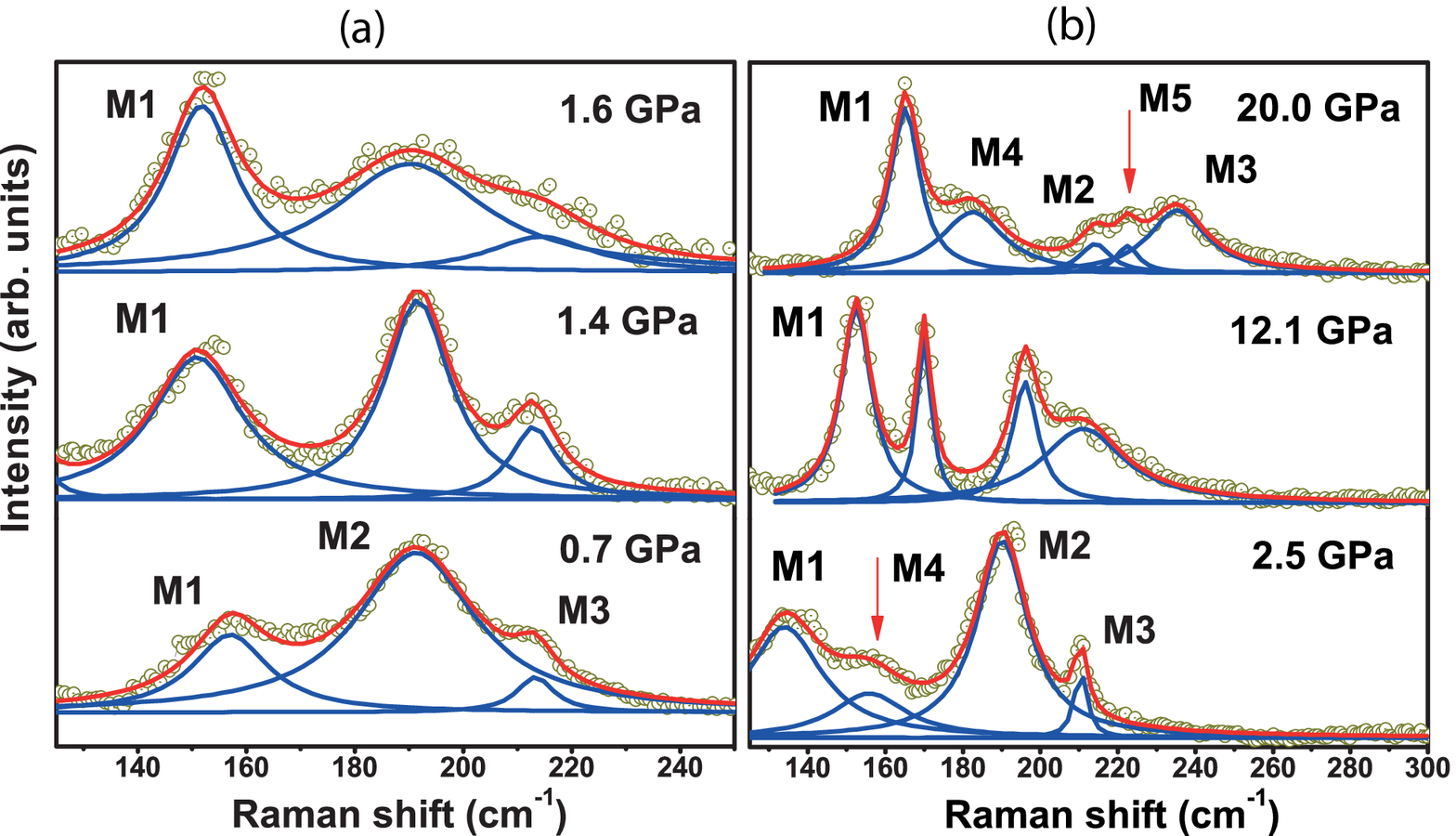}
\caption{\textit{(color online)--Pressure evolution of Raman spectra. The solid lines are Lorentzian fits to the experimental 
          data points. Appearance of new modes are indicated by arrows.}}
\label{Fig1}
\end{center}
\end{figure}

High pressure Raman experiments on single crystal of Sb$_2$Se$_3$ were carried out at room temperature (details in \cite{supp}). Group theory for a centrosymmetric rhombohedral structure of  Sb$_2$Se$_3$ $(D^5_{3d})$ predicts 12 optical zone center phonons represented by 2A$_{1g}$(R)+2E$_{g}$(R)+2A$_{2u}$(IR)+2E$_{u}$(IR) symmetry, where R and IR refer to Raman and infrared active modes respectively \cite{bs1,tex1}. Recently Raman studies have been pursued at high pressure to probe the reconstruction of Fermi surface topology giving rise to an ETT in TIs Bi$_2$Te$_3$, Bi$_2$Se$_3$ and Sb$_2$Te$_3$ \cite{poli1,bs0,gopal1,bs1,bs2}. Probing of ETT with pressure has been done either by observing changes in pressure coefficients ($\frac{d\omega}{dP}$), in full width at half maximum (FWHM) of Raman bands, or appearance of new modes, as observed in $\alpha$-Bi$_2$Te$_3$ \cite{bs0,gopal1}, $\alpha$-Bi$_2$Se$_3$ \cite{bs1} and $\alpha$-Sb$_2$Te$_3$ \cite{bs2}. It is clear from all the above mentioned studies that ETT occurs for all these TIs in $\sim$ 3 to 5 GPa hydrostatic pressure regime. Here, we report high pressure Raman studies of single crystal Sb$_2$Se$_3$ upto 24.6 GPa to show a ETT transition at $\sim$ 2.5 GPa marked by
(i) a large softening of the low frequency Raman mode by $\sim 16 \% $  together with an anomalous increase of its linewidth by $\sim 200 \% $ within a narrow pressure 
range of 0 to 2.5 GPa and relatively less softening by other two modes (M2 and M3), 
(ii) change in $\frac{d\omega}{dP}$ and
(iii) an appearance of a new mode.


\begin{figure}[p!]
\begin{center}
\leavevmode
\includegraphics[width=1.0\textwidth]{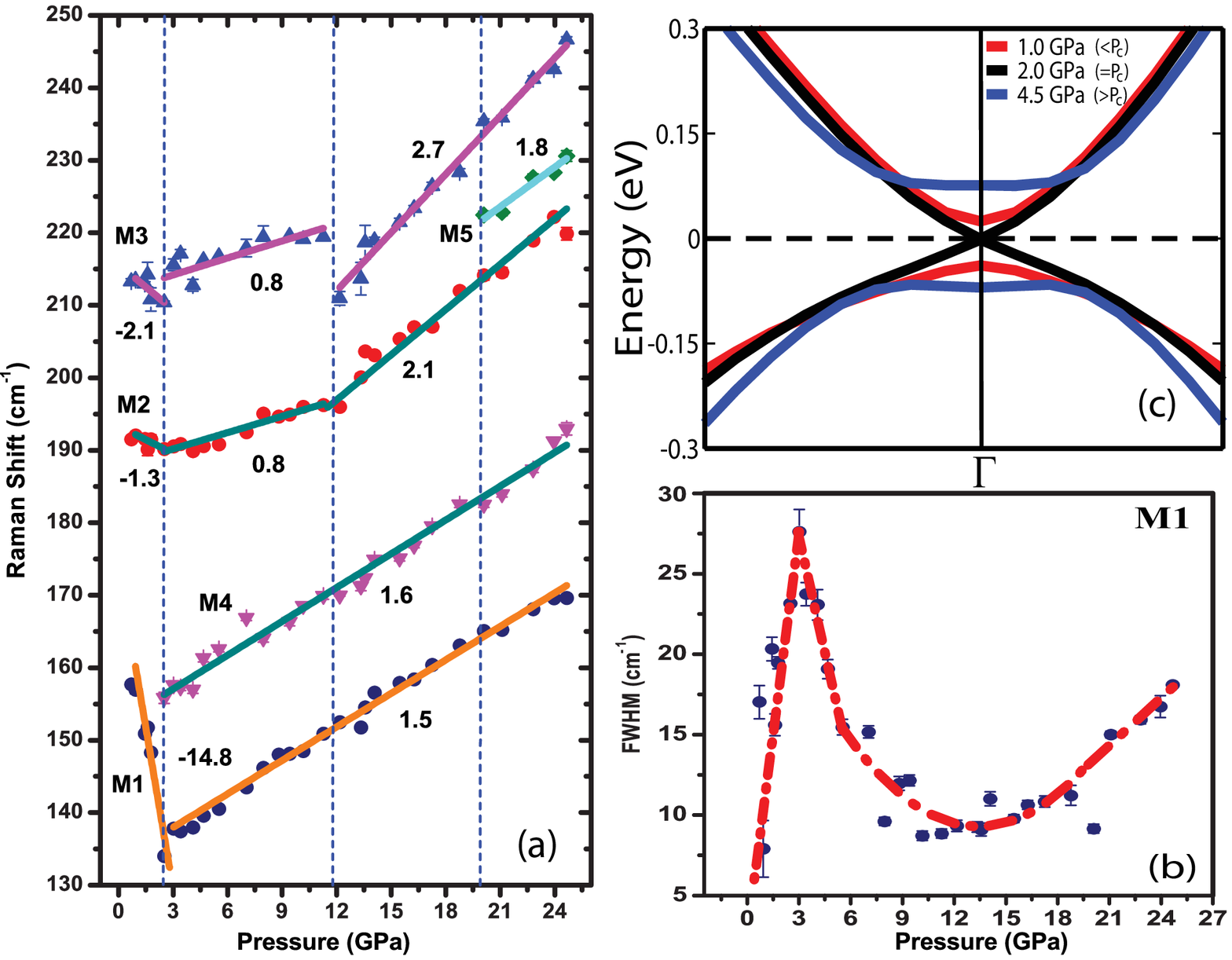}
\caption{\textit{(color online)--(a) Raman shift versus pressure plot. Solid lines are linear fits to the observed 
           frequencies. The numbers next to the straight lines are the fitted values of
					$\frac{d\omega}{dP}$ in cm$^{-1}$/GPa. The error bars, if not seen, are less than the size of the symbol. Fig.(b) shows FWHM of
					the M1 mode (solid points) as a function of pressure and the dashed line is drawn as guide to the eyes. Fig.(c) shows the First-principles calculations of electronic
					structure near the gap as a function of pressure in the neighborhood of transition ($P_c$=2 GPa).}}
\label{Fig2}
\end{center}
\end{figure}

Figure~\ref{Fig1} shows Raman spectra at a few pressures from 0.7 to 20.0 GPa. Three modes marked M1 ($\sim$ 166 cm$^{-1}$), M2 ($\sim$ 192 cm$^{-1}$) and M3 ($\sim$ 214 cm$^{-1}$) are clearly  seen at 0.7 GPa. As we shall see later, the mode M1 is assigned as  E$^2_g$, M2 as A$^2_{1g}$ and M3 as a combination mode. Mode frequencies extracted from Lorentzian fits to the spectra are plotted versus pressure in Fig.~\ref{Fig2}(a). The data  are fitted to the linear equation (solid line) $\omega_P=\omega_0+(\frac{d\omega}{dP})P$ in different high pressure regimes marked by vertical dotted lines in Fig.~\ref{Fig2}(a). The data in Fig. 2(a) have been shown to be reproducible in different runs.

The following observations from Fig.~\ref{Fig2}(a) and (b) are noteworthy: 
(i) M1 phonon mode shows a very large softening of $\sim$ 25 cm$^{-1}$ upto 2.5 GPa and then the sign of the 
slope ($\frac{d\omega}{dP}$) changes; whereas M2 and M3 show smaller softening but with a similar change of slope, 
(ii) The full width at half maximum (FWHM) of the M1 mode is maximum near 2.5 GPa and decreases on either side asymmetrically (Fig. 2(b)),
(iii) A new mode M4 appears at 2.5 GPa,
(iv) at $\sim$ 12 GPa, both M2 and M3 show a slope change,
(v) a new mode M5 appears around 20 GPa with frequency increasing linearly upto the observed maximum pressure (24.6 GPa). The emphasis of the present work is only on the  ETT transition at $\sim$ 2.5 GPa. A transition at low pressure around 3 GPa has been indicated in isostructural compounds Bi$_2$Te$_3$, Bi$_2$Se$_3$ and Sb$_2$Te$_3$ by change in pressure coefficients of the frequencies and their FWHM \cite{bs0, bs1, bs2}. In comparison, a large softening of phonons upto 2.5 GPa and anomalous linewidth maximum of the M1 mode as seen in this work is unique to Sb$_2$Se$_3$. Further, we clearly observe three phase transitions as shown in Fig.~\ref{Fig2}(a). Comparing our results with the previous 
X-ray and Raman studies under pressure on Bi$_2$Te$_3$ \cite{bs0}, Bi$_2$Se$_3$ \cite{bs1} and Sb$_2$Te$_3$  \cite{bs2}, the transition at $\sim$ 12 GPa can be associated with the rhombohedral to monoclinic C2/m phase and transition at $\sim$ 20 GPa to monoclinic C2/c structure. High pressure X-ray studies are required to confirm the high pressure phases reported here.


\begin{figure}[p!]
\begin{center}
\leavevmode
\includegraphics[width=1.0\textwidth]{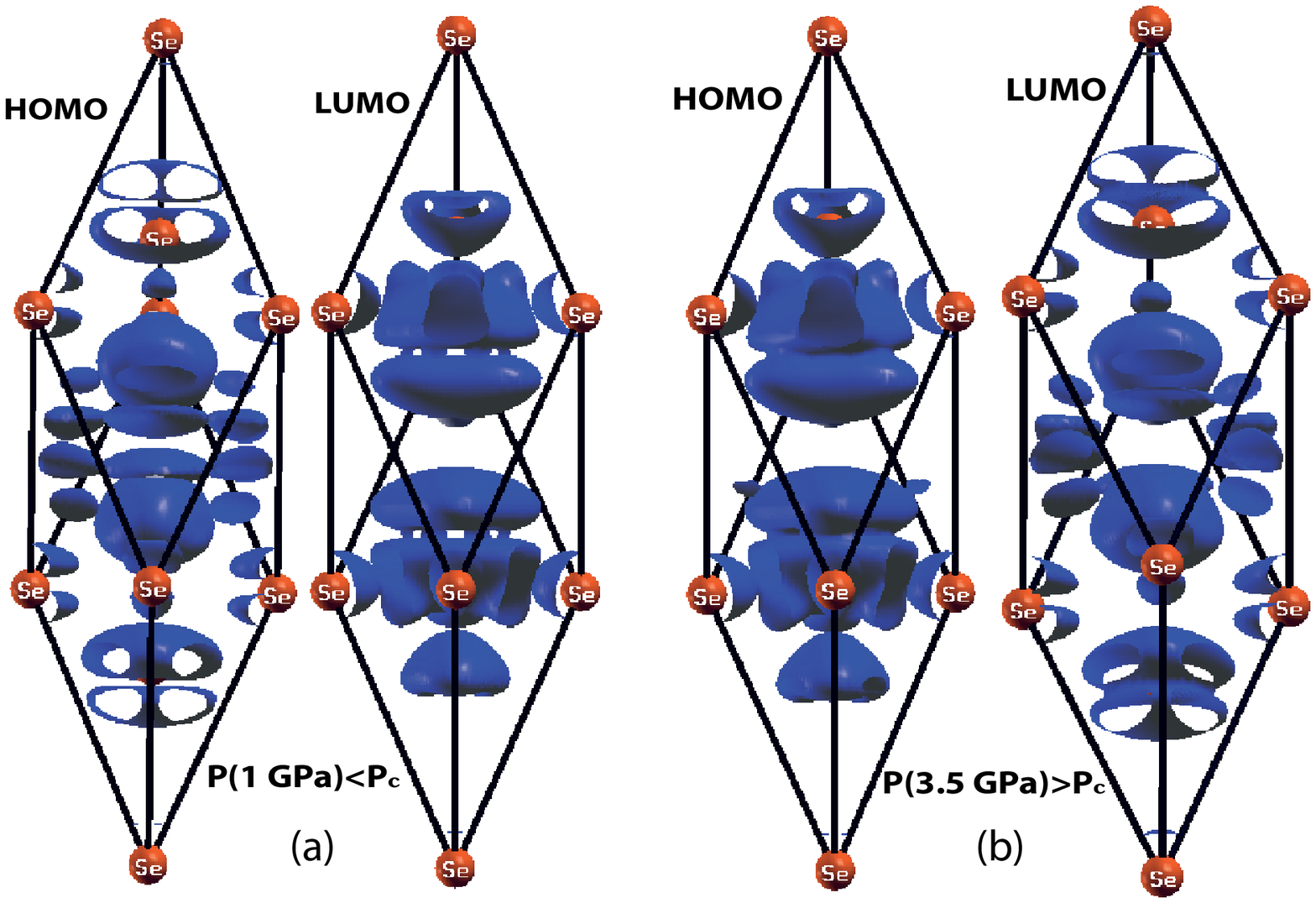}
\caption{\textit{(color online)-- First-principles evidence for the pressure induced electronic topological transition in Sb$_2$Se$_3$. Iso-surfaces of charge densities
  associated with electronic states at the top of valence and bottom of conduction bands at $\Gamma$ point (a) before and
 (b) after the ETT, demonstrating reversal of bands across the transition. Parity of these states is inferred from the specific Se atoms contributing to these states and also
 from visualization of various spinor components of these states (see \cite{supp}).}}
\label{Fig4}
\end{center}
\end{figure}

We now present results of first-principles DFT-based calculations of electronic structure and phonons \cite{supp}. For the rhombohedral structure of Sb$_2$Se$_3$ with $D^5_{3d}$ point group, we keep $c/a$ ratio fixed and examine properties as a function of pressure by varying the lattice constant  $a$ (Figs.~\ref{Fig2}(c) and 3). Our theoretical estimate of the lattice constant is $a=4.09 \AA$, at which a direct energy gap at $\Gamma$-point separates valence band states with odd parity from the conduction band states with even parity, as was shown by Zhang et al for a band insulator \cite{zhang1}. The odd parity of the state at the top of the valence band is evident in the involvement of the $p-$orbitals of Se atoms present at the inversion center. A strong hybridization between p-states of Sb and Se is evident in the electronic eigenstates across the gap (Fig.~\ref{Fig4}). Band gap vanishes precisely at $P_c$=2 GPa ($\textit{ Weyl metallic}$ $\it{state}$) and opens up for $P>P_c$ with energies varying linearly with $P$ near $P_c$. This involves a reversal of the parity of valence and conduction bands across $P_c$ marking a transition from a band insulator to a topological insulator \cite{naga1}.  Highly accurate calculations using two different techniques within adiabatic density functional theory -- linear response and frozen phonon methods -- yield phonon frequencies within a few cm$^{-1}$ of each other. However, calculated frequencies of all phonons at the $\Gamma$ point vary linearly with pressure, and exhibit  no anomaly at the transition at $P_c$ (Fig. S5 in \cite{supp}), leaving us puzzled about the mechanism of the sharp Raman anomalies. 

\begin{figure}[p!]
\begin{center}
\leavevmode
\includegraphics[width=1.0\textwidth]{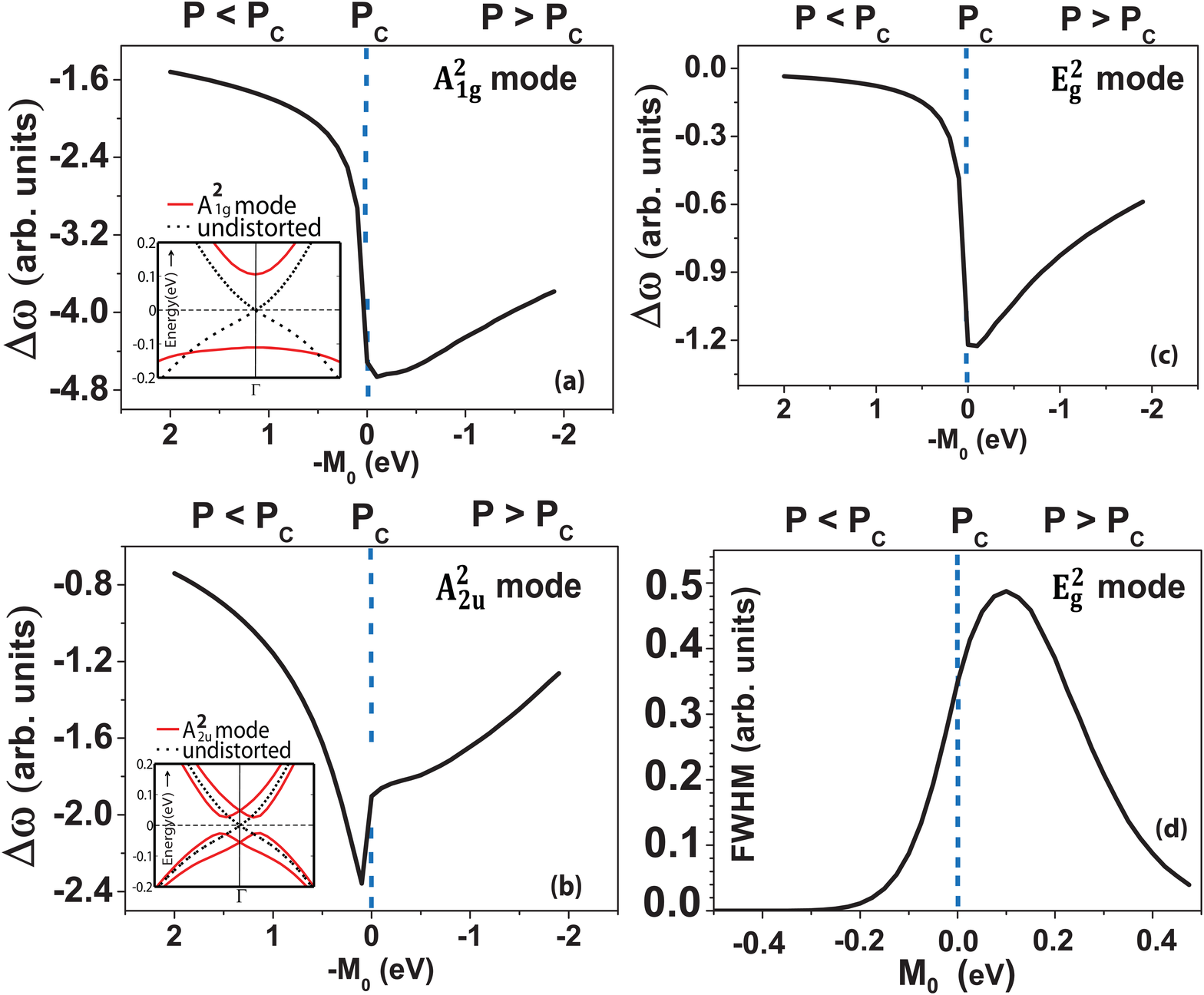}
\caption{\textit{(color online)-- Dynamical corrections to frequencies of phonon modes (a) $A^2_{1g}$, (b) $A^2_{2u}$ and (c) $E^2_{g}$ as a function of
 $-M_0$ $(i.e.\propto(P-P_c))$. The corrections are negative and make the modes softer near the transition; their asymmetry allows differentiation between the trivial and 
the nontrivial topology of electronic structure on the two sides of the transition. The insets in Fig. 4(a) and (b) show how electronic structure changes when atomic 
displacements of a given phonon mode are frozen to distort the structure based on first-principles calculations, and are reproduced using electron-phonon coupling (see text)
 within the 4-band model. (d) The calculated linewidth of the mode $E^2_{g}$ (M1 mode) as a function of $M_0$ i.e. pressure.}}
\label{Fig5}
\end{center}
\end{figure}

To this end, we now present theoretical analysis beyond the adiabatic approximation and determine dynamical \cite{ferrari1,andy1} corrections to phonon frequencies. We use the universal four band model  developed by Zhang et al \cite{model1,zhang1}  written in terms of Dirac matrices and note that states at the top of the valence and bottom of the conduction bands are doubly degenerate, and their energies are given by $C_0 \pm M_0$, where $C_0$ and $M_0$ are the parameters of the model Hamiltonian \cite{model1}, and $M_0<0$ for a topological insulator. Treating $M_0 = -\kappa (P-Pc)$ with a positive $\kappa$, we reproduce the electronic structure near the $\Gamma$ point close to the transition from a band insulator to a topological insulator (Fig.~\ref{Fig2}(c) and 3) as reflected in the reversal of bands of opposite parity \cite{supp}.

Electron-phonon coupling is needed to estimate dynamical corrections to vibrational frequencies, and we  now derive their form at the lowest order within the four-band model, expressing them in terms of Dirac matrices.  Symmetry properties of the Dirac matrices (irreducible representations (irreps) for the symmetry group of  Sb$_2$Se$_3$) have been derived by Liu et al \cite{model1}.  We note that Raman-active and IR-active modes  have even and odd parity, all of them are invariant under time reversal. An electron-phonon coupling term in the Hamiltonian can be expressed as a product of phonon degree of freedom and a Dirac matrix (or its commutator). A symmetry-invariant electron-phonon coupling is obtained by projecting this term onto identity representation of the double group of $D_{3d}^5$. For the Raman-active $A^2_{1g}$ mode, which has the full symmetry of the system, the electron-phonon coupling can have all the terms in the four-band electronic Hamiltonian multiplied by the mode displacement $u_{A_{1g}}$. We pick the leading term that gives changes in band structure obtained by distorting the structure with this mode (see inset of Fig.~\ref{Fig5}a): $H_{A_{1g}}=A_{A_{1g}}u_{A_{1g}}\Gamma_5,$ where $\Gamma_{5}$ is a Dirac matrix. Similarly, coupling of an IR-active mode $A^2_{2u}$ with electrons takes a form: $H_{A_{2u}}=A_{A_{2u}}u_{A_{2u}}\Gamma_{45},$	where $\Gamma_{ij}$ is a Dirac matrix commutator $[\Gamma_i , \Gamma_j]/2i$. Finally, symmetry consideration of the lowest order coupling of the $E^2_g$ mode with electrons requires one to have a quadratic form of Dirac matrices: $H_{E_{g}}=A_{E_{g}}(u^x_{E_{g}}\Gamma_{15}+u^y_{E_{g}}\Gamma_{25})\Gamma_{35},$	where $x$ and $y$ denote cartesian components of displacements of the doubly degenerate $E^2_g$ mode. In this term, each of matrix $\Gamma_{35}$ and the expression $u^x_{E_{g}}\Gamma_{15}+u^y_{E_{g}}\Gamma_{25}$ transform according to $\Gamma_1^{-}$ irrep and is even under time reversal, and their product is symmetry-invariant. Since the product of two non-commuting Hermitian operators is not Hermitian, we construct a symmetry-invariant Hermitian form: $H_{E_{g}}=A_{E_{g}} [(u^x_{E_{g}}\Gamma_{15}+u^y_{E_{g}}\Gamma_{25})\Gamma_{35} -h.c.]/2i,$ where h.c. means Hermitian conjugate \cite{tex2}.

We validate the form of electron-phonon coupling by comparing the electronic structure of the model with that obtained from first-principles for the lattice distorted with each mode of Sb$_2$Se$_3$ (insets of Figs.~\ref{Fig5}a and b for the case of $A^2_{2u}$ and $A^2_{1g}$ modes). For a frozen $A^2_{1g}$ mode at $P=P_c$, band gap opens up in a way similar to how it opens up at infinitesimally small deviation in pressure or strain. For a frozen $A^2_{2u}$ mode, band splitting is more interesting: each of the doubly degenerate valence and conduction bands of definite parity split up as this mode breaks the parity symmetry. The minima or maxima of bands shift away from the $\Gamma$ point. However, the splitting or changes in the model band structure associated with the electron-phonon  coupling of $E^2_g$ ($A_{E_g}$) mode are not captured within first-principles calculations,  suggesting that the  physics of this coupling is beyond the formal mean-field description of density functional theory used here. Thus, in the analysis below, we use the symmetry invariant form with $A$-coupling as a free parameter as the other terms are not needed to capture the essential aspects of the observed phonon anomalies.

We obtain the dynamical corrections to phonon frequencies as a function of pressure using these  forms of electron-phonon couplings (keeping strength of the coupling as free parameters) in first-order time dependent perturbation analysis \cite{ferrari1, supp} of the four-band model of Bi$_2$Se$_3$ derived by Zhang et al \cite{zhang1} with pressure  dependent $M_0$ (see Fig.~\ref{Fig5}). To simplify our analysis, we make use of the layered  nature of Sb$_2$Se$_3$ and carry out Brillouin zone integrations only in the ab-plane.  It is evident that dynamical corrections to frequencies of Raman active modes change sharply  below $P_c$ and asymmetrically around $P_c$.  Indeed, the sharp drop (Fig.~\ref{Fig5}) in frequencies of $E^2_g$ and $A^2_{1g}$ modes  just below $P_c$, and a gradual increase in their frequencies for $P>P_c$ are consistent with the pressure-dependent behavior of modes M1, M2 and M3 in Raman spectra reported here.  In contrast, dynamical corrections to IR-active mode exhibit a sharp jump above $P_c$!  Noting the calculated frequencies, we assign mode M1 to $E^2_g$ and mode M2 to $A^2_{1g}$ irreps.  The Mode M3 is the highest in frequency, and is most likely associated with the sum of two  $E^2_g$ modes (a second order Raman mode). Finally, the new mode M4 appearing above $P_c$  may correspond to  $A^2_{2u}$ (148 cm$^{-1}$), which becomes Raman active due to a change in the overall parity of the occupied states from even to odd. 

The linewidth of $E^2_g$ mode estimated from our analysis (Fig. 4d and Eq. 9 of \cite{supp}) peaks {\it asymmetrically} near the transition pressure, quite consistent with the observed linewidth anomaly seen in our  experiments (see Fig. 2(b)). This further corroborates our theoretical  analysis, and allows us to determine the origin of asymmetry in phonon anomalies to the parity reversal of occupied and unoccupied bands in immediate vicinity of the ETT. The similar analysis predicts the asymmetry in anomaly of IR active mode contrasting that in the Raman mode--with a jump in frequency above $P_c$. 

In conclusion, we reveal a pressure-induced electronic topological transition (ETT) in single  crystal of Sb$_2$Se$_3$, a band insulator. A combination of experiment, first-principles calculations and theoretical model-based analysis presented here show a breakdown of adiabatic approximation at the ETT. We established that electron-phonon coupling of nontrivial forms leads to anomalies in the observed  Raman spectra. These ideas are applicable to electronic transitions in other topological insulators too, and expected to stimulate experiments for exploring  anomalies in IR vibrational spectra, and guide materials scientists in transforming a normal insulating material to a topological insulator. 



AKS acknowledges the funding from Department of Science and Technology, India. UVW acknowledges funding from DAE-SRC Outstanding Researcher grant. PG thanks funding from AFOSR-MURI and an NSF Career award, DMR-0449969. AB thanks CSIR for a research fellowship. We thank Prof. M. Cardona for critical comments. \\



\begin{thebibliography}{29}
\bibitem{tex1}
The nominal frequencies for these modes, for example in Bi$_2$Se$_3$ are A$^1_{1g}$: 74 cm$^{-1}$, A$^2_{1g}$: 174 cm$^{-1}$, E$^1_g$: 42 cm$^{-1}$, E$^2_g$: 131 cm$^{-1}$.
\bibitem{tex2}
Similarly, there are other forms of electron and $E^2_g$ phonon couplings possible: $B_{E_{g}}[(u_{E_{g}}^{x}\Gamma_{24}-u_{E_{g}}^{y}\Gamma_{14})\Gamma_{34} - h.c.]/2i + B_{E_{g}}^{'}[(u_{E_{g}}^{x}\Gamma_{24}-u_{E_{g}}^{y}\Gamma_{14})\Gamma_{12} - h.c.]/2i$ \nonumber $+  C_{E_{g}}(k_{x}u_{E_{g}}^{x}-k_{y}u_{E_{g}}^{y})\Gamma_{4},$ where the term with coupling B and B$^{'}$ (C) involves products of $\Gamma_2^{+}$ ($\Gamma_2^{-}$) representations,  each of which is odd under time-reversal.
\end{thebibliography}
\end{document}